\documentclass[twocolumn,english,reprint, longbibliography, superscriptaddress, breaklinks=true, showkeys, showpacs=false, nofootinbib]{revtex4}
\usepackage[T1]{fontenc}
\usepackage[latin9]{inputenc}
\usepackage{color}
\usepackage{babel}
\usepackage{amsmath}
\usepackage{amssymb}
\usepackage{graphicx}
\usepackage{physics}
\usepackage[unicode=true,pdfusetitle,
 bookmarks=true,bookmarksnumbered=false,bookmarksopen=false,breaklinks=true,pdfborder={0 0 0},backref=false,colorlinks=true]
 {hyperref} 
\makeatletter
\@ifundefined{textcolor}{}
{%
 \definecolor{BLACK}{gray}{0}
 \definecolor{WHITE}{gray}{1}
 \definecolor{RED}{rgb}{1,0,0}
 \definecolor{GREEN}{rgb}{0,1,0}
 \definecolor{BLUE}{rgb}{0,0,1}
 \definecolor{CYAN}{cmyk}{1,0,0,0}
 \definecolor{MAGENTA}{cmyk}{0,1,0,0}
 \definecolor{YELLOW}{cmyk}{0,0,1,0}
}
\pdfoutput=1
\hypersetup{colorlinks=true,citecolor=blue,linkcolor=cyan,urlcolor=blue,filecolor= green, breaklinks=true}
\usepackage{url}
\usepackage{breakurl}
\makeatother

\begin{document}

\title{Complete complementarity relations: Connections with Einstein-Podolsky-Rosen realism and decoherence, and extension to mixed quantum states}

\author{Marcos L. W. Basso}
\email{marcoslwbasso@mail.ufsm.br}
\address{Departamento de F\'isica, Centro de Ci\^encias Naturais e Exatas, Universidade Federal de Santa Maria, Avenida Roraima 1000, Santa Maria, Rio Grande do Sul, 97105-900, Brazil}

\author{Jonas Maziero}
\email{jonas.maziero@ufsm.br}
\address{Departamento de F\'isica, Centro de Ci\^encias Naturais e Exatas, Universidade Federal de Santa Maria, Avenida Roraima 1000, Santa Maria, Rio Grande do Sul, 97105-900, Brazil}

\selectlanguage{english}%

\begin{abstract}
Wave-particle duality is an essential character of quantum systems. In the last few years, much progress has being made towards formally quantifying these quantum features. The properties of the quantum density matrix were shown to lead to duality inequalities for single-quanton mixed states and to triality identities, also known as complete complementarity relations (CCRs), for two-quanton pure states. 
In this article, we establish connections between CCRs and Einstein-Podolsky-Rosen (ir)realism, as formalized by Bilobran and Angelo in [Europhys. Lett. 112, 40005 (2015)].
Besides, we show that, for tripartite pure states $|\psi_{ABE}\rangle$, CCRs can be applied to obtain the system-environment entanglement of formation, $E_{f}(\rho_{AE})$, through the local information of a system $A$ and its classical correlation with an auxiliary system $B$. Moreover, we discuss CCRs in the context of the emergence of the pointer basis in the decoherence process  identified via the constancy of system-apparatus classical correlations. We also obtain CCRs for purified mixed bipartite quantum-classical states. 
In addition, we derive CCRs for mixed-bipartite two-qudit quantum states and discuss its interpretations.
\end{abstract}


\maketitle

\section{Introduction}
Bohr's complementarity principle was introduced as a qualitative statement about quantum systems, or quantons \cite{Leblond}, which have properties that are equally real, but mutually exclusive \cite{Bohr}. This principle, together with the uncertainty principle, is at the origin of Quantum Mechanics (QM), following the development of the theory since then. The wave-particle duality is the best known example of Bohr's principle, where, in a two-way interferometer, such as the Mach-Zehnder interferometer or the double-slit interferometer, the wave aspect is characterized by the visibility of interference fringes while the particle nature is given by the which-way information of the path along the interferometer. From an information-theoretic approach, Wootters and Zurek \cite{Wootters} were the first to investigate the wave-particle duality in a quantitative setting, looking at two-slit interference in the presence of a path detector they found  that simultaneous  observation  of  both complementary aspects is possible with the restriction that the more information it gives about one aspect of the system, the less information the experiment can provide about the other.  Later, Greenberger and Yasin \cite{Yasin} formulated a quantitative relation given by an inequality expressed in terms of a priori information about the path, named predictability, and fringe visibility. Englert \cite{Engle} took a different approach by considering a path quantifier which was based on a posteriori path information acquired using a path detector, and derived a stronger inequality between the distinguishability and the visibility. Until now, several different approaches were taken for quantifying the wave-particle properties of a quantum system \cite{Ribeiro, Bera, Coles, Hillery, Qureshi, Maziero, Lu}. For instance, recently it was realized that the quantum coherence \cite{Baumgratz} can be considered as a natural generalization for the visibility of an interference pattern \cite{Bera, Bagan, Mishra}.

It was shown in Ref. \cite{Maziero} that the positivity of the density matrix of a system $A$ leads to complementarity relations of the type
\begin{equation}
    C(\rho_{A})+P(\rho_{A})\le c(d_{A}), \label{eq:cr1}
\end{equation}
with $c(d_{A})$ depending only on the system $A$ dimension, where $C(\rho_{A})$ is a quantum coherence measure and $P(\rho_{A})$ is a corresponding predictability measure, with both satisfying the criteria established in Refs. \cite{Durr, Englert} for quantifiers of visibility and predictability. Besides, if we consider a purification of $\rho_{A}$, i.e., $|\psi\rangle_{AB}$ such that $\Tr_{B}(|\psi\rangle_{AB}\langle\psi|)=\rho_{A}$, with $\Tr_{B}$ being the partial trace function \cite{pTr}, then the complementarity relation above is completed \cite{Marcos, Leopoldo}:
\begin{align}
C(\rho_{A})+P(\rho_{A})+E(|\psi\rangle_{AB}) = c(d_{A}),  \label{eq:ccr1}  
\end{align}
with $E(|\psi_{AB}\rangle)$ being the entanglement \cite{Basso_ECCR} between $A$ and $B$. Triality relations like Eq. (\ref{eq:ccr1}) are also known as complete complementarity relations (CCRs), since in Ref. \cite{Qian} the authors interpreted this equality as completing the duality relation given by Eq. (\ref{eq:cr1}), thus turning the inequality into an equality. 

In view of these results, we are naturally compelled to investigate if and under what conditions it is possible to obtain a complete complementarity relation, 
\begin{align}
    C(\rho_{A})+P(\rho_{A})+K(\rho_{AB}) = c(d_{A}),
\end{align} 
with $Tr_{B}(\rho_{AB})=\rho_{A}$ and with $K(\rho_{AB})$ being a (quantum) correlation measure for the mixed state $\rho_{AB}$. This is one of the main objectives of this article. 

Besides, recently it was put forward an operational notion of (ir)realism by Bilobran and Angelo \cite{Renato}, who gave a formal and operational definition of the elements of reality first introduced by Einstein, Podolsky, and Rosen in Ref. \cite{Einstein}, where they stated that: \textit{``If, without in any way disturbing a system, we can predict with certainty (i.e., with probability equal to unity) the value of a physical quantity, then there exists an element of physical reality corresponding to this physical quantity.''} From this operational approach, Bilobran and Angelo were able to define a measure of (ir)realism of an observable given a preparation of a quantum system. In this article, we connect these two crucial notions at the heart of quantum mechanics by relating the measure of (ir)realism with complementarity relations, since it is expected that an element of reality of a given path (or, more generally, an arbitrary observable) in an interferometer is related to the capability of predicting with probability equal to one the path of the quanton a priori, or if, in principle, it is possible to access a posteriori the information about the path in some other degree of freedom.

The remainder of this article is organized in the following manner. In Sec. \ref{sec:realism}, we establish connections between CCRs and (ir)realism. In Sec. \ref{sec:purifications}, we explore CCRs for bipartite mixed states that are obtained from tripartite purifications. Moreover, In Sec. \ref{sec:mixed}, we obtain CCRs for mixed bipartite quantum states, and discuss its interpretations. Finally, in Sec. \ref{sec:conc}, we give our conclusions.

\section{Complementarity and its connection with (ir)realism}
\label{sec:realism}
For any quantum state $\rho_A$ of dimension $d_A$, the relative entropy of coherence is defined as  \cite{Baumgratz} 
$
    C_{re}(\rho_A) = \min_{\iota \in I} S_{vn}(\rho_A||\iota),
$
where $I$ is the set of all incoherent states, which are those states which are diagonal in a reference basis, and $S_{vn}(\rho_A||\iota) = \Tr(\rho_A \log_2 \rho_A - \rho_A \log_2 \iota)$ is the relative entropy. The minimization procedure leads to $\iota = \rho_{Adiag} = \sum_{i = 1}^{d_A} \rho^A_{ii} \ketbra{i}$. Thus 
\begin{align}
    C_{re}(\rho_A) = S_{vn}(\rho_{Adiag}) - S_{vn}(\rho_A) \label{eq:cre}.
\end{align}
Once $C_{re}(\rho_A) \le S_{vn}(\rho_{Adiag})$, it is possible to obtain an incomplete complementarity relation from this inequality:
\begin{equation}
    C_{re}(\rho_A) + P_{vn}(\rho_A) \le \log_2 d_A \label{eq:cr6},
\end{equation}
with $P_{vn}(\rho_A) := \log_2 d_A - S_{vn}(\rho_{Adiag}) = \log_2 d_A + \sum_{i = 0}^{d_A - 1} \rho^A_{ii} \log_2 \rho^A_{ii}$ being a measure of predictability, already defined in Ref. \cite{Maziero}. We observe that it is  possible to define this predictability measure once the diagonal elements of $\rho_A$ can be interpreted as a probability distribution, which is a consequence of the properties of $\rho_A$. Now, if $\rho_A$  can be regarded as a subsystem of a bipartite pure quantum system $\ket{\Psi}_{AB}$, then it is possible to assign the incompleteness of the complementarity relation (\ref{eq:cr6}) to the presence of correlations and take $S_{vn}(\rho_A)$ as a measure of entanglement between the subsystems $A$ with $B$. So, it is possible to interpret Eq. (\ref{eq:cre}) as a complete complementarity relation
\begin{equation}
C_{re}(\rho_A) + P_{vn}(\rho_A) + S_{vn}(\rho_A) = \log d_A \label{eq:ccre},
\end{equation}
as already obtained in  \cite{Marcos}. It is worth mentioning that Eq. (\ref{eq:ccre}) is equivalent to that obtained in Ref. \cite{Ribeiro} using a different reasoning.

Besides, from Eq. (\ref{eq:ccre}), here we connect complementarity  with the operational notion of (ir)realism put forward quite recently in Ref. \cite{Renato}. First, we consider a preparation $\rho_A$ of the quantum system $A$. Second, it is performed, between the preparation and the tomography procedures, a non-selective projective measurement of an observable $\mathcal{O}$ (which can be the path of a Mach-Zehnder interferometer). Here $\mathcal{O} = \sum_k o_k \Pi^{\mathcal{O}}_k$  is considered to be a discrete spectrum observable,  with  $\Pi^{\mathcal{O}}_k = \ketbra{o_k}{o_k}$ being orthonormal projectors acting  on $\mathcal{H}_A$. In addition, in this section, the coherence and the predictability measures are taken with regard the eigenbasis of the observable  $\mathcal{O}$. Since any information about the measurement outcomes is not revealed, the post measurement state is given by
\begin{align}
    \Phi_{\mathcal{O}}(\rho_{A}) =  \sum_k  \Pi^{\mathcal{O}}_k \rho_{A} \Pi^{\mathcal{O}}_k.
\end{align}
Therefore, when $\Phi_{\mathcal{O}}(\rho_{A}) = \rho_{A}$, the observer can conclude that an element of reality for $\mathcal{O}$ was already implied in the preparation. Thus, the authors in Ref. \cite{Renato} took the process of non-revealed measurements $\Phi_{\mathcal{O}}(\rho_{A})$ as  the main element of the reality of the observable $\mathcal{O}$ given the preparation $\rho_{A}$ and defined the following measure of local irreality (or indefiniteness) of $\mathcal{O}$ given the preparation $\rho_{A}$: 
\begin{equation}
\mathfrak{I}(\mathcal{O}|\rho_{A}):= S_{vn}(\Phi_{\mathcal{O}}(\rho_{A})) - S_{vn}(\rho_{A}).
\end{equation}
It is worth mentioning that, in Ref. \cite{Renato}, the authors introduced the notion of irrealism of an observable $\mathcal{O}$ for bipartite quantum systems, as well. Here, we shall deal only with local irreality. Besides, this measure was already applied in several investigations by Angelo and co-authors \cite{Dieguez, Gomes, Angelo, Fucci, Orthey, Moreira}.

Now, it is possible to define the local reality (or definiteness) of the observable $\mathcal{O}$ given the local state $\rho_A$ as
\begin{align}
    \mathfrak{R}(\mathcal{O}|\rho_A) & := \log_2 d_A - \mathfrak{I}(\mathcal{O}|\rho_A) \nonumber \\ 
    &  = \log_2 d_A + S_{vn}(\rho_{A}) - S_{vn}(\Phi_{\mathcal{O}}(\rho_{A})). 
\end{align}
Given this definition of local reality of the observable $\mathcal{O}$, it is straightforward to connect it with complementarity. First, let us consider that $\rho_A$  can be regarded as a subsystem of a bipartite pure quantum system $\ket{\Psi}_{AB}$, as before. By noticing that
\begin{align}
    \Phi_{\mathcal{O}}(\rho_{A}) = \sum_k \Pi^{\mathcal{O}}_k  \rho_{A} \Pi^{\mathcal{O}}_k = \rho_{Adiag},
\end{align}
i.e., $\Phi_{\mathcal{O}}(\rho_{A})$ is diagonal in the eigenbasis of the observable $\mathcal{O}$, then it is straightforward to see that
\begin{align}
     \mathfrak{R}(\mathcal{O}|\rho_A) + C_{re}(\rho_A) = \log_2 d_A \label{eq:rea},
\end{align}
which is equivalent to Eq. (\ref{eq:ccre}). Indeed,
\begin{align}
    \mathfrak{R}(\mathcal{O}|\rho_A) & =  \ln d_A - S_{vn}(\rho_{Adiag}) + S_{vn}(\rho_A) \\
    & = P_{vn}(\rho_A) + S_{vn}(\rho_A),
\end{align}
which expresses the fact that the local reality of the path observable $\mathcal{O}$ is related to the predictability of the observable $\mathcal{O}$ before a projective measurement, i.e., its ``pre-existing'' reality as well as the possible generation of entanglement with an informer, i.e., a degree of freedom that records the information about the state of the system. Besides, it is noteworthy that 
\begin{equation}
C_{re}(\rho_A) =  \mathfrak{I}(\mathcal{O}|\rho_A),
\end{equation}
which means that the local irreality of the observable $\mathcal{O}$ is directly related to the quantum coherence of $\rho_A$ in the eigenbasis of $\mathcal{O}$, or, equivalently, to the wave aspect of the quanton $A$ in respect to the observable $\mathcal{O}$, as already shown in Ref. \cite{Renato}. 
Thus, the results reported in this section formally connect elements of (ir)reality with complementarity.

\section{Tripartite purifications of bipartite mixed states}
\label{sec:purifications}
It was shown in Ref. \cite{Koashi} that for a pure tripartite state $|\psi_{ABE}\rangle$, it follows that $$S_{vn}(\rho_{A})=E_{f}(\rho_{AB})+J_{A|E}(\rho_{AE}),$$ where $J_{A|E}(\rho_{AE})=S_{vn}(\rho_{A})-\min_{\{\Pi_{j}^{E}\}}\sum_{j}p_{j}S(\rho_{j}^{A})$ is a measure of classical correlation of the system $A$ with the environment $E$ (or with just another auxiliary quantum system). Above $p_{j}=\Tr_{A}(\Pi_{j}^{E}\rho_{AE}\Pi_{j}^{E})$ and $$\rho_{j}^{A}=Tr_{E}(\Pi_{j}^{E}\rho_{AE}\Pi_{j}^{E})/p_{j},$$ where $\Pi_{j}^{E}$ are the elements of a positive operator value measurement (POVM). Also, for $\rho_{AB}=\sum_{j}p_{j}|\psi_{j}^{AB}\rangle\langle\psi_{j}^{AB}|$, the entanglement of formation is defined as $$E_{f}(\rho_{AB})=\min_{\{p_{j},|\psi_{j}^{AB}\rangle\}}\sum_{j}p_{j}E_{E}(|\psi_{j}^{AB}\rangle),$$ where the entanglement entropy is $E_{E}(|\psi_{j}^{AB}\rangle)=S_{vn}(\Tr_{B}(|\psi_{j}^{AB}\rangle\langle\psi_{j}^{AB}|))$. In this way, the CCR for relative entropy, Eq. (\ref{eq:ccre}), can be rewritten as
\begin{equation}
    E_{f}(\rho_{AB})+J_{A|E}(\rho_{AE})+P_{vn}(\rho_{A})+C_{re}(\rho_{A})=\log_{2}(d_{A}). \label{eq:entan}
\end{equation}
Therefore, one can see that the local reality of the observable $\mathcal{O}$, given the state $\rho_A$, depends on the entanglement of formation between $A$ and $B$, the classical correlations between $A$ and $E$ and the predictability of $A$. Besides, it is possible to interchange the roles of $B$ and $E$ such that the entanglement of formation between the  system $A$ and the environment, $E_{f}(\rho_{AE})$, can be inferred via the local information about $A$, i.e., $P_{vn}(\rho_A)$ and $C_{re}(\rho_A)$, and the classical correlations between $A$ and $B$, $J_{A|B}(\rho_{AB})$, where $B = \mathcal{A}$ can be e.g. the apparatus $\mathcal{A}$, that is measuring an observable of the system $A$, or any auxiliary system.  This is a possible application of complete complementarity relation, since the detection of entanglement between a system $A$ and the environment $E$ is receiving a lot of a attention recently \cite{Roszak, Katarzyna}.

In addition, it is noteworthy that in Ref. \cite{Cornelio} the authors showed that, for a system subject to the process of decoherence, the pointer basis emerges when the classical correlation  between system and apparatus, $J_{A|\mathcal{A}}(\rho_{A\mathcal{A}})$, becomes  constant, even though system $+$ apparatus still have quantum features, i.e., $\rho_{A \mathcal{A}}$ is not a classical state. We see from Eq. (\ref{eq:entan}), $E_{f}(\rho_{AE})+J_{A|\mathcal{A}}(\rho_{A\mathcal{A}})+P_{vn}(\rho_{A})+C_{re}(\rho_{A})=\log_{2}(d_{A})$, that, after the emergence of the pointer basis, changes in the local properties of the system $A$ are due to its entanglement with the environment $E$. For instance, for the dephasing channel \cite{Wilde}, for which $P_{vn}(\rho_{A})$ is constant, the rate of decrease of the quantum coherence is equal to the rate of entanglement creation, i.e., $-\partial_{t}C_{re}(\rho_{A})=\partial_{t}E_{f}(\rho_{AE}),$ with $\partial_{t}\equiv\frac{\partial}{\partial t}.$

Also related to the emergence of the pointer basis in the decoherence process, in Ref. \cite{Dieguez} the authors used the operational definition of (ir)realism to discuss the measurement problem. They considered the scenario proposed by Everett \cite{Everett}, in which an external observer $O_e$ describes a measurement conducted within a laboratory by an internal observer $O_i$, and it was shown that the amount of information acquired by the internal observable about the system $A$ through the apparatus $\mathcal{A}$  quantified by the conditional information
\begin{align}
    \mathcal{I}_{A|\mathcal{A}} := \log_2 d_A - S_{A|\mathcal{A}}(\rho_{A \mathcal{A}})
\end{align}
is always the same regardless of the reference frame that one chooses to assess it, where $S_{A|\mathcal{A}}(\rho_{A \mathcal{A}}) = S_{vn}(\rho_{A \mathcal{A}}) - S_{vn}(\rho_{\mathcal{A}})$ is conditional quantum entropy \cite{Wilde}. In $O_i$'s frame, it is used the notion of state collapse and the average information over individual runs of the experiment, whereas in $O_e$'s frame the same informational content is obtained by considering a unitary evolution plus the discard of $O_i$. Here we relate the works of Ref. \cite{Cornelio} and Ref. \cite{Dieguez}. Both the classical correlation $J_{A|\mathcal{A}}$ and the conditional information $\mathcal{I}_{A|\mathcal{A}}$ are constant after the emergence of the pointer basis through the decoherence process, even though the definitions of $J_{A|\mathcal{A}}$ and $\mathcal{I}_{A|\mathcal{A}}$ are different. In fact, it is possible to show that the conditional information can be rewritten as $\mathcal{I}_{A|\mathcal{A}} = \ln d_A - \sum_j p_j S(\rho_{A|\Pi_j^{\mathcal{A}}})$ \cite{Dieguez}. Similarly to $J_{A|\mathcal{A}}$, it's possible maximize $\mathcal{I}_{A|\mathcal{A}}$ over all projective measures $\{\Pi^{\mathcal{A}}_j\}$ to identify the pointer basis. Thus, $\mathcal{I}_{A|\mathcal{A}} \ge J_{A|\mathcal{A}}.$ However, the main message in both works is the same: any measurement must be repeatable and verifiable by other observers. According to the Copenhagen interpretation, reductions of the wave packet must  return  the  same  information  at any  time  and  by  any  observer.

Finally, let us regard \textbf{quantum-classical states}. It is interesting to notice that, given the purification $|\psi_{ABE}\rangle$, by considering unrevealed projective measurements $\{\Pi^B_j \}$ on the apparatus $B = \mathcal{A}$, the quantum-classical state\footnote{One possible purification for these states is $|\Psi\rangle_{ABE}=\sum_{j,k}\sqrt{p_{j}a_{jk}}|a_{jk}\rangle_{A}\otimes|j\rangle_{B}\otimes|c_{j,k}\rangle_{E}$, with $|c_{j,k}\rangle_{E}$ being an orthonormal basis for $\mathcal{H}_{E}$ (i.e., $\langle c_{jk}|c_{j'k'}\rangle=\delta_{j,j'}\delta_{k,k'}$ and $\sum_{j,k}|c_{j,k}\rangle\langle c_{j,k}|=\mathbb{I}_{E}$) and $\rho_{A|\Pi^{B}_j}=\sum_{k=1}^{d_{A}}a_{jk}|a_{jk}\rangle\langle a_{jk}|.$} $\rho_{AB} = \sum_j p_j \rho_{A|\Pi^B_j} \otimes \Pi^B_j$ satisfies the complete complementarity relation (cf. Eq. (\ref{eq:ccre}))
\begin{align}
    P_{vn}(\rho_{AB}) + C_{re}(\rho_{AB}) + S_{vn}(\rho_{AB}) = \log_{2}(d_A d_B),
    \label{eq:ccrAB}
\end{align}
with $S_{vn}(\rho_{AB}) $ measuring the entanglement between $AB$ and $E$. Now
\begin{align}
    S_{vn}(\rho_{AB}) & = S_{vn}\Big(\sum_j p_j \rho_{A|\Pi^B_j} \otimes \Pi^B_j\Big) \nonumber \\
    & = H(p_j) + \sum_j p_j S_{vn}(\rho_{A|\Pi^B_j}), \label{eq:ccre1}
\end{align}
where $ H(p_j) = - \sum_j p_j \log_{2}(p_j)$ is Shannon's entropy. Analogously, $S_{vn}(\rho_{{AB}_{diag}}) = H(p_j) + \sum_j p_j S_{vn}((\rho_{A|\Pi^B_j})_{diag})$. Hence
\begin{align}
      P_{vn}(\rho_{AB}) & = \log_{2}(d_A d_B) - S_{vn}(\rho_{{AB}_{diag}})  \\ 
      & = \log_{2}(d_B) - H(p_j) - \sum_j p_j P_{vn}(\rho_{A|\Pi^B_j}),\nonumber \\
     C_{re}(\rho_{AB}) & = S_{vn}(\rho_{{AB}_{diag}}) - S_{vn}(\rho_{AB}) \nonumber \\
     & = \sum_j p_j C_{re}(\rho_{A|\Pi^B_j}).
\end{align}
Thus, using these relations, it follows from Eq. (\ref{eq:ccrAB}) that
\begin{align}
    0 & = \sum_j p_j \Big(P_{vn}(\rho_{A|\Pi^B_j}) + C_{re}(\rho_{A|\Pi^B_j}) \nonumber \\ 
    & \hspace{1.5cm} + S_{vn}(\rho_{A|\Pi^B_j}) - \log_{2}(d_A) \Big), 
\end{align}
i.e., each member of the ensemble $\{p_j, \rho_{A|\Pi^B_j}\}$ satisfies a complete complementarity relation:
\begin{align}
    P_{vn}(\rho_{A|\Pi^B_j}) + S_{vn}(\rho_{A|\Pi^B_j}) + C_{re}(\rho_{A|\Pi^B_j})  = \log_{2}(d_A).
\end{align}
So, in this case we obtain a CCR in the ``standard form'', with $S_{vn}(\rho_{A|\Pi^B_j})$ measuring the entanglement between $A$ and $E$, when we fix $j$.

\section{Complete complementarity relations for mixed states}
\label{sec:mixed}

A natural question that arises is the following one: given a bipartite quantum system, if $\rho_{AB}$ is not pure, can Eqs. of the type (\ref{eq:ccre}) still completely quantify the complementarity behavior of the quanton $A$? The first thing to notice here is that in this case $S_{vn}(\rho_A)$ cannot be considered as a measure of entanglement between the subsystem $A$ and $B$. In fact, in this case, $S_{vn}(\rho_A)$ is just a measure of the mixedness (or the uncertainty) of the quanton $A$. 

In the context of mixed states, Tessier \cite{Tessier} obtained the following complementarity relation for two qubits
\begin{equation}
    \Tr \rho \tilde{\rho} + S_l({\rho_{AB}}) + \Bar{S}^2(\rho_A) + \Bar{S}^2(\rho_B) = 1,
    \label{eq:tessier}
\end{equation}
where $\Tr \rho \tilde{\rho} = 1 - \Tr \rho^2_A - \Tr \rho^2_B + \Tr \rho^2_{AB}$ with $\tilde{\rho} = \sigma_y \otimes \sigma_y \rho^*_{A,B} \sigma_y \otimes \sigma_y$ is a measure of multipartite entanglement for qubits introduced in Ref. \cite{Jaeger}. Besides, $S_l({\rho_{AB}})=1-\Tr\rho_{AB}^{2}$ is the linear entropy of $\rho_{AB}$, which can be considered as a measure of the mixedness of the bipartite quantum system, whereas $\Bar{S}^2(\rho_k) = \frac{1}{2}(V^2(\rho_k) + P^2(\rho_k)), \ \ k = A,B$, are the averages of the squares of the single qubit properties, i.e., its visibility and predictability, as defined originally in Ref. \cite{Yasin}. Here, by applying the definitions of $C$, $P$, and $\mathcal{I}_{A:B}$, we obtain an informational complementarity relation, analogous to Eq. (\ref{eq:tessier}), for two qudit states:
\begin{align}
    \log_{2}(d_A d_B) & =  \mathcal{I}_{A:B}(\rho_{AB}) + S_{vn}(\rho_{AB}) \nonumber  \\ 
    & \hspace{0.3cm}+ \sum_{k = A, B}\Big(P_{vn}(\rho_k)  + C_{re}(\rho_k)\Big)  \label{eq:info},
\end{align}
where $S_{vn}(\rho_{A,B})$ is measuring the mixedness of the whole system. By noticing that $I(\rho_{AB}) := \log_2(d_A d_B) - S_{vn}(\rho_{A,B})$ is a measure of the state information of the system $AB$, as defined in Ref. \cite{Costa}, it is straightforward to see that
\begin{align}
    I(\rho_{AB}) = \mathcal{I}_{A:B}(\rho_{AB}) + \sum_{k = A, B}\Big(P_{vn}(\rho_k) + C_{re}(\rho_k)\Big),
\end{align}
which tell us that the information contained in the state $\rho_{AB}$ of the bipartite quantum system is given by the local wave-particle aspects of $A$ and $B$ and by their mutual information $\mathcal{I}_{A:B}(\rho_{AB})$. In fact, as showed in Ref. \cite{Costa}
$
    I(\rho_{AB}) = I(\rho_A) + I(\rho_B) + \mathcal{I}_{A:B}(\rho_{AB}), 
$
where $I(\rho_k) := \ln d_k - S_{vn}(\rho_k)$ is local information of the subsystem $k = A,B$. Therefore, we can see that the predictability and visibility of the quanton correspond to the local information contained in the reduced density matrix, since $$I(\rho_k) = P_{vn}(\rho_k) + C_{re}(\rho_k)$$ for $k = A,B.$ 

Besides, it is easy to see that Eq. (\ref{eq:info}) can be rewritten as
\begin{align}
\log_{2}(d_A) &= \mathcal{I}_{A:B}(\rho_{AB}) + S_{A|B}(\rho_{AB}) \nonumber \\ 
& \hspace{0.5cm} + P_{vn}(\rho_A) + C_{re}(\rho_A), \label{eq:cond}
\end{align}
where $ S_{A|B}(\rho_{AB}) = S_{vn}(\rho_{AB}) - S_{vn}(\rho_B)$. One can see that this CCR constrains the local aspects of the quanton $A$ by its correlations with the quanton $B$ given by $\mathcal{I}_{A:B}(\rho_{AB})$ and by the remaining ignorance about $A$ given that we have access to the system $B$. However, it is worth mentioning that the quantum conditional entropy can be negative, even though the sum $\mathcal{I}_{A:B}(\rho_{AB}) + S_{A|B}(\rho_{AB})$ is always positive. It is also noteworthy that, in this case, the local reality of the observable $\mathcal{O}$ is given by $ \mathfrak{R}(\mathcal{O}|\rho_A) = \mathcal{I}_{A:B}(\rho_{AB}) + S_{A|B}(\rho_{AB}) + P_{vn}(\rho_A).$ Besides, from Ref. \cite{Wilde},  we have the following definition of coherent information: $S_{A>B}(\rho_{AB}) := -S_{A|B}(\rho_{AB}) = S(\rho_B) - S(\rho_{A,B}),$ which can be interpreted as an information quantity that  \textit{is measuring the extent to which we know less about part of the a system than we do about its whole}. For instance, if we have a maximally pure entangled state, then $S_{A>B}(\rho_{AB}) = S_{vn}(\rho_B) = \log_{2}2$, which means that we know more about the whole than about its parts. Motivated by this interpretation and by the protocol of state merging \cite{Horodecki}, we consider the following interpretation for the quantum conditional entropy in the context of complementarity relations: $S_{A|B}(\rho_{AB})$ is an information quantity that \textit{is measuring the extent in which it is more worth it to know about the system $A$ than about the whole system}.

For instance, let us consider the following situations:
\begin{enumerate}
    \item If $\rho_{AB}$ is a maximally entangled pure state of two qubits, then $S_{A|B}(\rho_{AB}) = - \log_2 2$. Therefore, it is more worth to know about the whole system than  about $A$, since we have an entangled state which allows one to implement quantum protocols (like state merging, teleportation, etc).
    \item $\rho_{AB} = \rho_A \otimes \rho_B \implies  S_{A|B}(\rho_{AB}) = S_{vn}(\rho_A) > 0$. So, it is more worth it to know about $A$ than about the whole system, once we need more qubits of information to know about the whole system than about $A$. In other words, besides to having knowledge about $A$, it is necessary to know about $B$. Beyond that, in this case $\mathcal{I}_{A:B}(\rho_{AB}) = 0 $ and the local part of the complementarity relation $S_{A|B}(\rho_{AB})+ P_{vn}(\rho_A) + C_{re}(\rho_A) = \log_{2}(d_A)$ gives information if we can use the system for classical or quantum protocols.  If our system has quantum coherence, we can use it to execute some quantum protocols, meanwhile if $P_{vn}(\rho_A)$ is maximum, we can address a classical bit to the state of the system.
    \item $\rho_{AB} = \frac{1}{4} I_{4 \times 4} \implies S_{A|B}(\rho_{AB}) = S_{vn}(\rho_A) = \log_2 2$. Even though it is useless to know about $A$ (which is in an incoherent state), it costs two times less to know that $A$ is an incoherent state than to know that $AB$ is an incoherent state.
\end{enumerate}

Finally, Eq.(\ref{eq:cond}) can also be rewritten as
\begin{align}
    \mathcal{I}_{A|B}(\rho_{AB}) =  \mathcal{I}_{A:B}(\rho_{AB}) + P_{vn}(\rho_A) + C_{re}(\rho_A),
\end{align}
where $\mathcal{I}_{A|B}(\rho_{AB}) = \log_2 d_A - S_{A|B}(\rho_{AB})$ is conditional information defined in Ref. \cite{Dieguez}, and which gives the information content about $A$ that can be accessed from $B$. In contrast with the conditional quantum entropy, the conditional information is always positive, since $\mathcal{I}_{A|B}(\rho_{AB}) \ge \mathcal{I}_{A:B}(\rho_{AB}) \ge 0$. It is noteworthy that the left-hand side refers to the part $B$, while the right-hand side gives us information about the local properties of $A$ and about its correlations with $B$.


\section{Conclusions}
\label{sec:conc}
Wave-particle duality is one of the most fascinating and fundamental aspects of quantum theory. Recently, several investigations have been addressed towards formally quantifying the wave-particle aspects of quantum systems. In particular, it was shown that the positivity and unit trace of the quantum density operator leads to duality inequalities for an one-quanton mixed state and to triality equalities for two-quanton pure states. Continuing with this line of research, in this article we established connections between complete complementarity relations (CCRs) and EPR (ir)realism, an important advance connecting two concepts of fundamental importance for the foundations of Quantum Mechanics. Besides, considering tripartite purifications of bipartite density operators, we showed how CCRs can be used to quantify the quanton-environment entanglement via the quanton's local properties and its classical correlation with an auxiliary quantum system. 
In addition, we also discussed CCRs in the context of the emergence of the pointer basis in the decoherence process identified via the constancy of system-apparatus classical correlations. As well, we obtained CCRs for purified mixed-bipartite quantum-classical states.
At last, we obtained CCRs for mixed two-qudit states and explored its interpretations.

\begin{acknowledgments}
This work was supported by the Coordena\c{c}\~ao de Aperfei\c{c}oamento de Pessoal de N\'ivel Superior (CAPES), process 88882.427924/2019-01, and by the Instituto Nacional de Ci\^encia e Tecnologia de Informa\c{c}\~ao Qu\^antica (INCT-IQ), process 465469/2014-0.
\end{acknowledgments}

\end{document}